\documentclass[10pt,aps,twocolumn,prb,floatfix,superscriptaddress,
showpacs,footinbib]{revtex4-2}

\usepackage{amsmath}
\usepackage{mathptmx}
\usepackage{newtxmath}
\usepackage[T1]{fontenc}
\usepackage[latin9]{inputenc}
\usepackage{xcolor}
\usepackage{verbatim}
\usepackage{float}
\usepackage{graphicx,float}
\usepackage{silence}
\usepackage[percent]{overpic}

\usepackage{graphicx}% Include figure files
\usepackage{blindtext, xcolor}
\usepackage{xcolor}
\usepackage{hyperref}
\hypersetup{
	colorlinks,%
	citecolor=blue,%
	linkcolor=blue,%
	urlcolor=blue
}

\begin{document}
	\title{Magnetic response and antiferromagnetic correlations in  strained kagome ribbons}
	
	\author{R. F. P. Costa}
	
	\affiliation{Instituto de F\'isica,  Universidade Federal  de Uberl\^andia,Uberl\^andia, MG 38400-902, Brazil}
	
	\author{E. Vernek}
	\affiliation{Instituto de F\'isica,  Universidade Federal  de
		Uberl\^andia,Uberl\^andia, MG 38400-902, Brazil}
	
	\affiliation{Department of Physics and Astronomy, and Nanoscale \&
		Quantum Phenomena Institute, Ohio University, Athens, Ohio 45701-2979, USA}
	
	\author{S. E. Ulloa}
	\affiliation{Department of Physics and Astronomy, and Nanoscale \&
		Quantum Phenomena Institute, Ohio University, Athens, Ohio 45701-2979, USA}
	
	\begin{abstract}
		We study the physics of the strong-coupling Hubbard model in a kagome lattice ribbon under mechanical tension and half-filling. It is known that in the absence of strain, the lattice symmetry of the system and strong electronic interactions induce magnetic frustration. As uniaxial strain is applied, the ribbon exhibits various configurations with energy oscillations that depend on the direction of the strain axis. The ground states are obtained by density-matrix renormalization-group calculations. We find that the system is characterized by strong antiferromagnetic bonds distributed throughout the lattice in directions and patterns that depend on strain directions and may coexist with easily polarizable sites that are only weakly correlated to their neighbors. We identify frustration and correlation measures that follow the strain and interaction dependence of the system well.
		These results illustrate that strain-dependent magnetic susceptibility could be explored experimentally to help probe the role of symmetry and interactions in these systems.
	\end{abstract}
	\maketitle
	
	\section{Introduction}
	
	The quantum spin liquid is an exotic and fascinating phase of matter~\cite{RevModPhys.89.025003,Broholm2020,Kivelson2023}. Proposed by Anderson in~\cite{Anderson1973}, he  suggested that such a phase could emerge as the ground state of a spin-1/2 antiferromagnet on geometrically frustrated triangular lattices~\cite{Fazekas1974}. However, after years of debate on whether a quantum spin-liquid (QSL) phase would emerge in a triangular antiferromagnet, detailed studies suggest this is not the case (see \cite{RevModPhys.89.025003} and references therein).  Indeed, the search for the realization of this elusive phase of matter has elicited great effort from the scientific community over the years~\cite{RevModPhys.88.041002,PhysRevX.14.021010,Starykh2024,Wen2019,Imai2016}.

	In this context, the triangular motives present in kagome materials have attracted much attention in recent years~\cite{Wang2023}.  Their unique crystal structure, consisting of corner-shared triangles arranged around a honeycomb structure, induces high geometrical frustration,  offering thus the tantalizing possibility of stabilizing a QSL ground state~\cite{PhysRevLett.81.2356,PhysRevB.76.064430,yan_spin_liquid_2011}. 
	Although magnetic insulating kagome lattices have been investigated for a long time~\cite{Syozi1951}, the interest in their electronic and magnetic properties has motivated a reexamination in the context of QSL phases. Theoretical studies on kagome lattices have explored interesting physical phenomena, including magnetic~\cite{PhysRevB.45.2899,PhysRevB.78.174420,PhysRevLett.101.117203,Nakano2011}, and electronic~\cite{PhysRevB.90.081105,PhysRevB.80.113102,PhysRevB.99.165141,okamoto_topological_2022,Mojarro2023} properties, QSL phases~\cite{yan_spin_liquid_2011,Peng2021,PhysRevLett.119.067002}, and  unconventional superconductivity~\cite{PhysRevLett.110.126405,Kang2022,Mielke2022,PhysRevB.106.174514,PhysRevB.107.184106,PhysRevLett.127.177001}. Several materials have also been investigated experimentally, prominently including FeSn~\cite{Kang2019,Li2022}, CoSn~\cite{Kang2020}, MnSn~\cite{Park2018} and the rich class of materials AV$_3$Sb$_5$ with (A=K,Rb,Cs)~\cite{PhysRevMaterials.3.094407}, and RV$_6$Sn$_6$ with (R=Gd, Ho,Y)~\cite{Gong2015,Yin2021}.

	Different phases have been predicted theoretically in frustrated kagome lattices using antiferromagnetic Heisenberg and/or Hubbard models~\cite{PhysRevB.100.060408}. While the former accounts strictly for the spin degrees of freedom in the system, the latter also accounts for charge fluctuations and can then describe a wider range of phenomena.  Schnyder et al.~\cite{PhysRevB.78.174420} showed that different ground states could be obtained by controlling the spatial anisotropy of antiferromagnetic Heisenberg couplings on a kagome lattice, resulting in spiral phases and possible noncoplanar order. More recently, Nayga and Vojta~\cite{PhysRevB.105.094426} have also shown that the ground state of a kagome magnet with classical spins can be controlled by applied mechanical distortion of the lattice, including the finding that triaxial strain results in a noncoplanar spin-liquid phase. These studies point out that the magnetic properties of kagome systems are very sensitive to changes in the geometric frustration of the lattice.
	The reduction in symmetry would then be expected to strongly affect systems that incorporate charge fluctuations, such as those present in kagome metals.  The  Hubbard model provides a natural description, as the relative hoppings can be seen as effective control parameters that allow one to address the important question of how strain modifies the ground state properties.  Sun and Zhu~\cite{PhysRevB.104.L121118} have shown recently that by controlling the Coulomb repulsion in a Hubbard model of a kagome ribbon it is possible to drive the system across multiple quantum phase transitions. This includes a strongly interacting regime where the ground state is effectively governed by an antiferromagnetic Heisenberg model and displays QSL characteristics. 
	For weaker interaction, these authors saw evidence of different correlated phases in this geometrically frustrated system.  We are interested in exploring the role that strain plays in the behavior of kagome systems as interactions change their response from the metallic to the insulating regime. The question we want to explore is how strain modifies possible frustrated ground states in the different regimes with varying correlations and charge fluctuations.
	
	In this work, we investigate the effect of a uniaxial uniform
	strain in a kagome nanoribbon. A schematic representation of the system is depicted in Fig.~\ref{schematic_ribbon}. Adopting a single-band  Hubbard model, the effective role of strain is to modify the hopping matrix elements between electronic orbitals localized at the lattice sites. This generally results in anisotropic couplings along the three lattice vectors, thus reducing geometric frustration in the system.
	Furthermore, varying the Coulomb repulsion allows glancing at how magnetic properties of the metallic and insulating ground states differ, thereby highlighting the relevance of charge fluctuations.

	By performing density-matrix renormalization-group (DMRG) calculations within a tensor networks platform~\cite{ITensor,ITensor-r0.3}, we compute the
	ground state properties of the system, including the spatial distribution of spin-spin correlations and the local magnetization
	upon application of weak magnetic fields. 
	Our results show that the anisotropy produced by the applied strain induces the formation of Neel lines of strongly antiferromagnetically correlated
	bonds arranged on different patterns and along different lattice
	directions, depending on the strain orientation. This behavior
	can be understood as arising from the suppression of the geometrical frustration in the system.
	Interestingly, we also find that scattered among the well-defined Neel lines with antiferromagnetic correlations, sites exist that are only weakly coupled to their neighbors. These interstitial sites are thus easily polarizable by weak external magnetic fields, while the Neel lines remain in their antiferromagnetic coupled structure. The pattern of Neel bonds and spatial distribution of polarizable sites depend critically on strain orientation and becomes more pronounced for larger values of Coulomb interaction as the system approaches the kagome antiferromagnet regime.
	To better characterize the response to strain, we introduce two intuitive quantities: a geometric indicator to quantify the structural frustration in the system and another that measures how local spin correlations reflect the degree of frustration in the ground state. These measures are good predictors of the different spin correlation profiles seen in the system as the strain and interaction change.
	
	The remainder of this paper is organized as follows: In
	Sec.~\ref{model_and_methods} we introduce the model and methods to study
	the kagome lattice. In Sec.~\ref{results_section} we present and discuss our main findings. Finally, our concluding remarks are presented in Sec.~\ref{conclusions}.
	
	\section{Model and numerical Methods}
	\label{model_and_methods}
	\begin{figure}[b!]
		
		\includegraphics[scale=0.58, trim = 0 0.25cm 0 0.5cm]{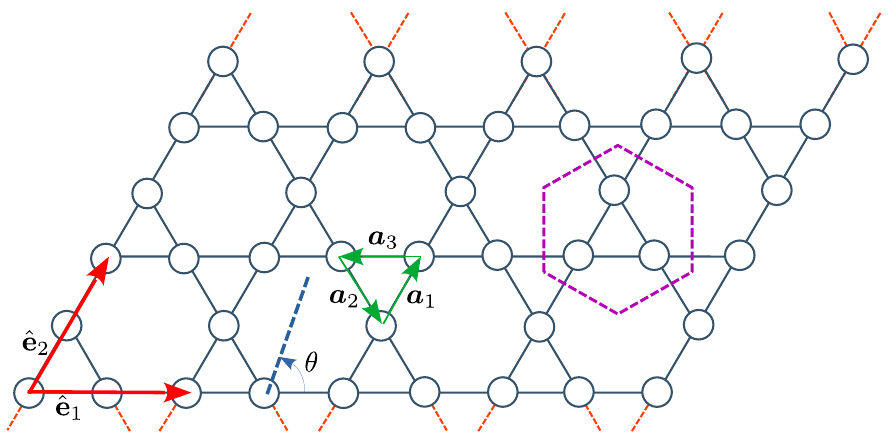}
		\caption{Schematic representation of the kagome nanoribbon corresponding
			to $L_x = 4$ unit cells in the $x$ direction and $L_y$ = 3 unit
			cells in the $\hat{\bf e}_2$ direction. Circles represent the atomic sites of a tight-binding model, while solid lines correspond to nearest-neighbor hopping matrix elements. Red dashed lines at the top and bottom
			indicate periodic boundaries, in which case the ribbon folds into
			what will be referred to as the long cylinder geometry. Directions $\hat{\bf e}_1$ and $\hat{\bf e}_2$ span the unit cell, while the ${\boldsymbol a}_j$ ($j=1,2,3$) represent the nearest-neighbor vectors. The angle $\theta$ defines the direction of applied
			strain with respect to $\hat{\bf e}_1$. The dashed magenta hexagon contains a
			minimal unit cell and associated bonds used in the text.
			\label{schematic_ribbon}}
	\end{figure}
	
	We consider a kagome lattice ribbon (see Fig.~\ref{schematic_ribbon})
	described by a tight-binding  Hubbard Hamiltonian as
	\begin{equation}\label{model}
		\begin{aligned}
			H = \varepsilon_0\sum_{j,\sigma}n_{j\sigma} - \sum_{\substack{<i,j> \ \sigma}}t_{ij}c_{i\sigma}^{\dagger}c_{j\sigma} + \sum_{j} \left(Un_{j\uparrow}n_{j\downarrow}+V_zS^{z}_{j}\right),
		\end{aligned}
	\end{equation}
	where $c_{j\sigma}^\dagger$ ($c_{j\sigma}$) creates (annihilates) an
	electron with energy $\varepsilon_0$ and spin $\sigma$ at site $j$, 
	$n_{j\sigma}=c_{j\sigma}^\dagger c_{j\sigma}$ is the number operator and
	$S^z_{j}=\hbar\left(n_{j\uparrow}- n_{j\downarrow}\right)/2$ is the local spin $z$-component. 
	$U$ represents the on-site Coulomb repulsion, and $V_z=g\mu_B B$ the Zeeman energy due to a magnetic field $B$ in the $z$-direction. 
	{As defined in Appendix~\ref{appendix_strained_couplings}, hopping matrix elements between nearest-neighbor sites}
	(denoted by $<i,j>$), 
	are modified by the presence of a spatially
	homogeneous uniaxial strain in the ribbon as
	\begin{eqnarray}
		t_{ij}\equiv t_\alpha\left(\theta\right)=t\exp\left[-\beta\left(\left\Vert
		\boldsymbol{a}^\prime_{\alpha}\left(\theta\right)\right\Vert
		-1\right)\right],\label{eq:strained_hoppings_main_text}
	\end{eqnarray}
	where $\beta$ is the material-dependent Gr\"uneisen parameter ($\simeq 3$ \cite{GrueneisenParam}),
	$\boldsymbol{a}^\prime_{\alpha}=\left(\boldsymbol{I}+\overline{\epsilon}\right)
	\boldsymbol{a}_{\alpha}$ are the strained vectors, $\boldsymbol{a}_{1}=\frac{1}{2}
	\left(1,\sqrt{3}\right),$ $\boldsymbol{a}_{2}=\frac{1}{2}
	\left(1,-\sqrt{3}\right)$ and $\boldsymbol{a}_{3}=-(\boldsymbol{a}_{1}+\boldsymbol{a}_{2})=\left(-1,0\right)$
	define 
	the unstrained vectors (see Fig.~\ref{schematic_ribbon}), and set the nearest neighbor separation as the unit length.
	The strain tensor is \cite{Mojarro2023}
	\begin{equation}
		\overline{\epsilon}=\epsilon\left[\begin{array}{cc}
			\cos^{2}\theta-\nu\sin^{2}\theta & \left(1+\nu\right)\sin\theta\cos\theta\\
			\left(1+\nu\right)\sin\theta\cos\theta & \sin^{2}\theta-\nu\cos^{2}\theta
		\end{array}\right],\label{eq:strain-tensor-main-text}
	\end{equation}
	where, $\epsilon$ is the strain strength, $\nu$ the material's
	Poisson ratio and $\theta$ the direction
	of applied strain, as indicated in Fig.~\ref{schematic_ribbon}.
	
	We are interested in the low-energy physics of the system, so that our analysis involves mainly the ground state and a few excited ones, as needed. We employ the density matrix renormalization group (DMRG) approach implemented within the iTensor library \cite{ITensor,ITensor-r0.3}, a matrix product states platform suitable for obtaining many relevant physical quantities in quantum many-body Hamiltonians. We obtain associated physical quantities characteristic of the system, such as local magnetizations $M^\nu_j=\langle S^\nu_j \rangle$ and spin-spin correlations $C_{ij}=\langle {\boldsymbol S}_i\cdot {\boldsymbol S}_j \rangle - \langle {\boldsymbol S}_i \rangle \cdot \langle {\boldsymbol S}_j \rangle$, where ${\bf S}_j$  is the spin operator on a given lattice site $j$, and $\langle \cdots\rangle$ represents the expectation value in the low-energy manifold. As defined, $C_{ij}$ gives the intrinsic many-body correlations in the presence of an applied magnetic field by removing the trivial contribution induced by the polarizing field. In the absence of a field, $M^\nu_j=0$ for all $\nu$, so that $C_{ij}=\langle {\boldsymbol S}_i\cdot {\boldsymbol S}_j \rangle$ reduces to the usual expression.  For visualization, it is useful to define the spatial link correlation as
	\begin{eqnarray}\label{link_corr}
		{\cal C_{\text{link}}}(x,y)=\sum_{i,j} {C}_{ij}  e^{-d_{ij}(x,y)/b}, 
	\end{eqnarray}
	where $d_{ij}(x,y)$ is the distance between any point on the 
	$xy$-plane and the segment that connects the lattice sites $i$ and $j$, while
	$b$ controls the sharpness of the link structure. 
	
	\section{Numerical Results}
	\label{results_section} 
	We set $t=1$ as the energy unit and $\hbar=1$ hereafter, and ensure that the system is in the half-filling regime by setting $\varepsilon_0=-U/2$. We explore values of $U$ from $U=5$ to $U=20$, corresponding to different phases of the system. For large values of $U$, the system is well described by an antiferromagnetic Heisenberg model \cite{PhysRevB.104.L121118}. The number of sites in the kagome nanoribbon increases rapidly with $L_x$ and $L_y$, $N=L_y(3L_x+2)$;
	we focus our analysis on a system with $L_x=12$ and $L_y=3$. {Whenever the ribbon is folded along the $\boldsymbol{\hat{\text{e}}_2}$ direction into a cylinder, with sites periodically coupled as indicated by red dashed lines in Fig.~\ref{schematic_ribbon}, we call such arrangement the \textit{long cylinder geometry (LCG)}, which} corresponds to the YC6 cylinder in Ref.~\cite{yan_spin_liquid_2011}. On completely open boundaries, however, the configuration will be referred to as the \emph{flat ribbon geometry (FRG)}. We have allowed DMRG link dimension up to 5000, resulting in truncation errors typically less than 10$^{-6}$.
	
	\subsection{Unstrained vs strained regimes} 
	\label{no_strain_subsection}
\begin{figure}[h!]
		\hskip -0.35cm 	
		\includegraphics[scale=0.495]{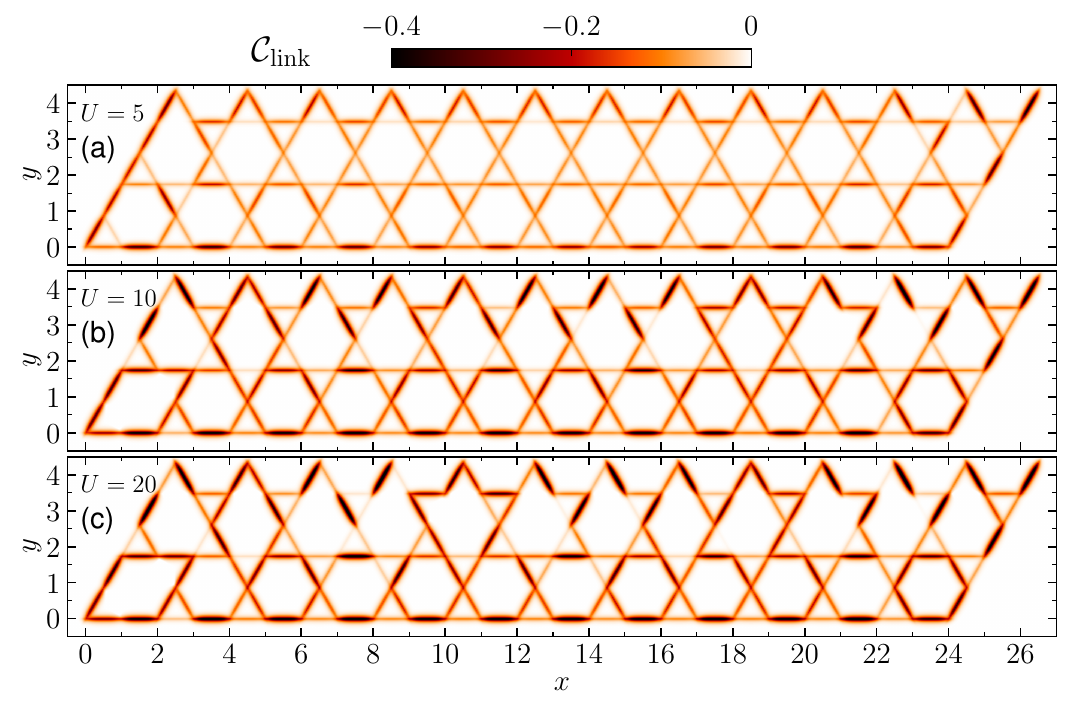}
		\caption{Link correlations, ${\cal C}_{\rm link}$, in the absence of strain ($\epsilon=0$) for  (a) $U=5$, (b) $U=10$ and (c) $U=20$. Darker links represent stronger antiferromagnetic alignments, clearly enhanced in panel (c). The uniform link distribution in (a) becomes more inhomogeneous for increasing $U$. Apart from strong links along the edges of this flat ribbon, there appears to be no overall order of the antiferromagnetic bonds in (c).
			\label{link_frustrated}}
	\end{figure}

	We first look at the spin-spin correlations in the system for a strain-free FRG lattice, $\epsilon=0$. As defined above, we use ${\cal C}_{\rm link}$ to provide a convenient visualization of the correlations.  Figure~\ref{link_frustrated} shows a heat map of ${\cal C}_{\rm link}$ for different Hubbard interaction values, placing the system into different phases: $U=5$ (a), metallic phase; $U=10$ (b) and $U=20$ (c), quantum spin liquid/kagome antiferromagnetic phase \cite{PhysRevB.104.L121118}. As the color bar shows, darker links correspond to stronger antiferromagnetic pairs. For small $U$, the correlation links are not as strong and are homogeneously distributed, as the weak interactions in this metallic regime would suggest. Despite the homogeneous distribution of hopping terms in the Hamiltonian, we notice some differences, likely due to edge effects on the ribbon.  For larger $U$, the antiferromagnetic correlations 
	appear stronger (darker links) and not as homogeneously distributed as the system approaches the Heisenberg antiferromagnetic regime.  In an infinite system, the spin-dominated ground state is expected to have high frustration
	without strain. A spin-liquid phase may even emerge, with not yet fully understood characteristics \cite{yan_spin_liquid_2011,PhysRevB.104.L121118}. We aim to study the role that strain may play in modifying the frustration otherwise present in the system. We will contrast Fig.~\ref{link_frustrated} with the link correlation in the presence of strain to expose some of its effects on the ground state. All link correlations plots in this paper are for FRG.

	\begin{figure}[h!]
		\includegraphics[scale=0.51]{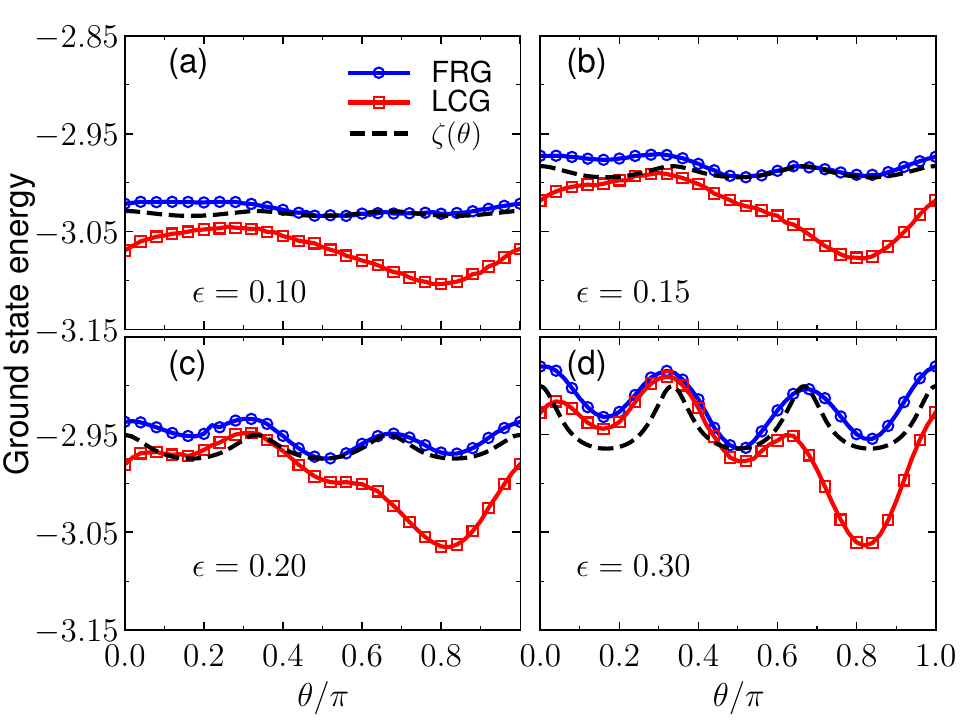}
		\caption{(a) Ground state energy per site, $E_0/N$, as  function of strain direction $\theta$ for $\epsilon=0.1$ (a), $\epsilon=0.15$ (b), $\epsilon=0.2$ (c) and $\epsilon=0.3$ (d). Blue and red curves correspond to Flat Ribbon Geometry and Long Cylinder Geometry respectively. The black dashed line shows $\zeta(\theta)$ defined in 
			Eq.~\eqref{coupling_isotropicity} and scaled as per~\cite{Note1}. In all panels $U=5$, i.e., in the metallic regime for the unstrained system.
			\label{coupling_isotropicity_vs_GS_energies}}
	\end{figure}
	\begin{figure*}
		\centering
		\includegraphics[width=\textwidth]{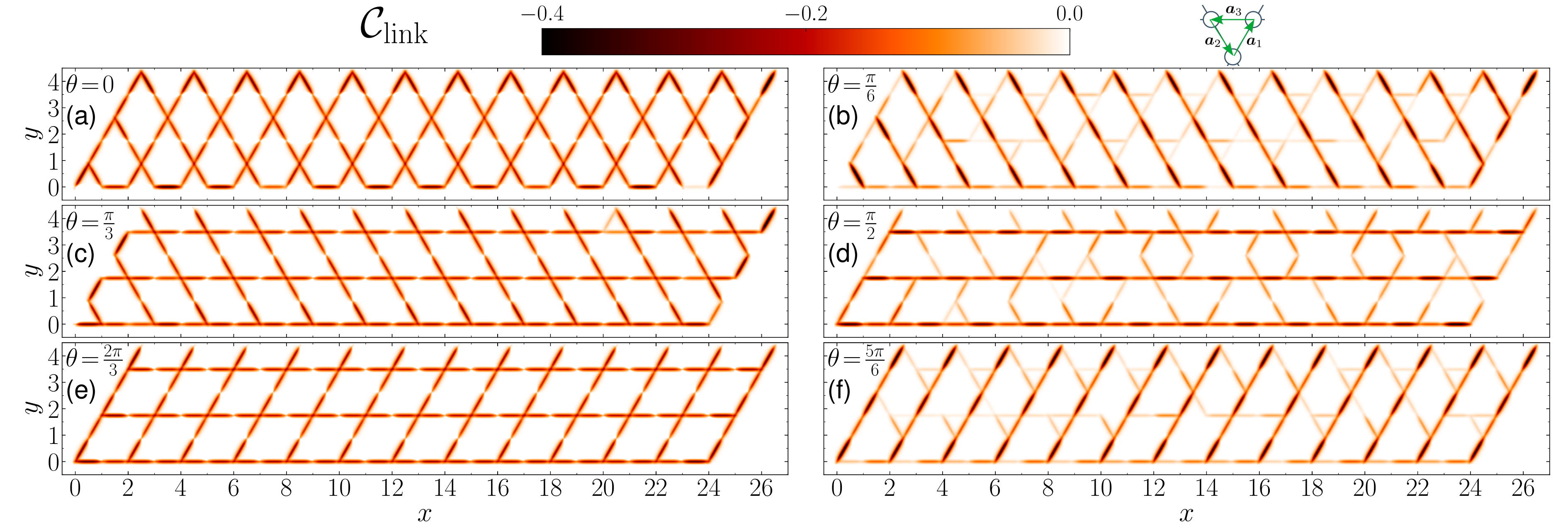}
		\caption{Link correlations for $\epsilon=0.30$ and different strain orientations: $\theta=0$ (a), $\theta=\pi/6$ (b), $\theta=\pi/3$ (c), 
			$\theta=\pi/2$ (d), $\theta=2\pi/3$ (e) and $\theta=5\pi/6$ (f). 
			The left (right) column panels correspond to \emph{maxima (minima) in $E_0(\theta)$ of Fig.~\ref{coupling_isotropicity_vs_GS_energies}.}
			For \emph{energy maxima (left column)}, links with antiferromagnetic correlations form 2D oblique Lieb-like lattices.
			For \emph{energy minima (right column)}, links form 1D-like arrays with stronger (darker) antiferromagnetic correlations. $U=5$ in all panels. The top-right shows an inset with nearest-neighbor vectors to help identify the link orientations.}
		\label{link_array}
	\end{figure*} 
	\begin{figure*}
		\centering
		\includegraphics[width=\textwidth]{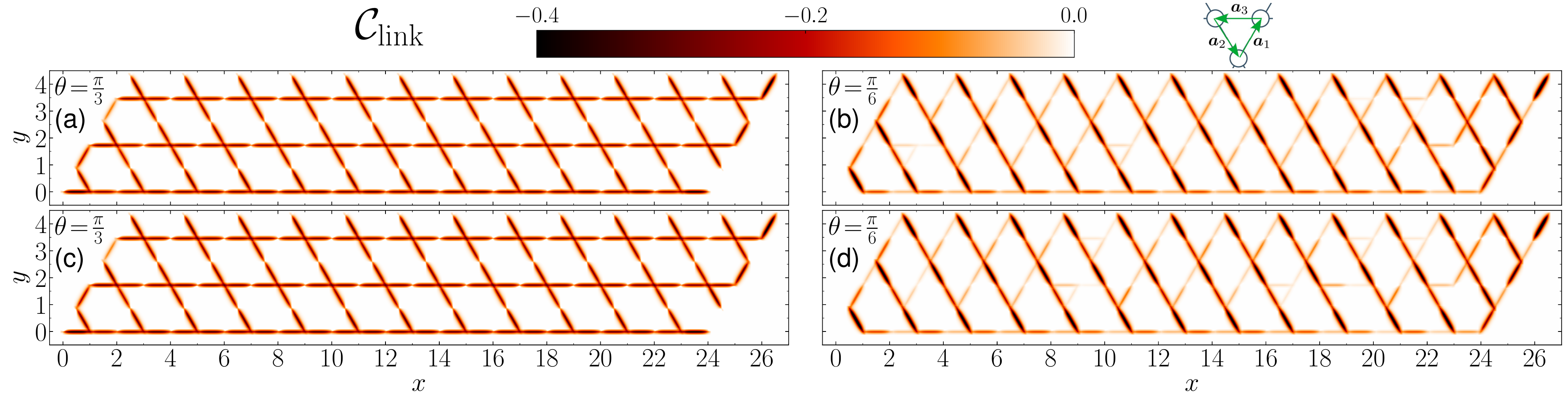}
		\caption{Link correlations for $\epsilon=0.30$ along $\boldsymbol{a_1}$ ($\theta=\pi/3$, left panels, \emph{a maximum in $E_0(\theta)$}) and orthogonally to $\boldsymbol{a_2}$ ($\theta=\pi/6$, left panels, \emph{a minimum in $E_0(\theta)$}). These configurations occur whenever strain is applied along $\boldsymbol{a_1}$ and orthogonally to ${\boldsymbol{a_2}}$, in this order (see top-right inset). Upper and lower panel rows correspond to $U=10$ and $U=20$, respectively. Compared to Fig.~\ref{link_array}(b) and (c), the correlations here are much more pronounced, with little difference between $U=10$ and 20.}
		\label{link_array_larger_U}
	\end{figure*}
	We now analyze the effect of strain on the link-correlation pattern over the ribbon. Because the effect of strain on the correlations depends on the angle $\theta$, it is instructive to first analyze how the ground state energy $E_0$ changes with $\theta$. Figure~\ref{coupling_isotropicity_vs_GS_energies} shows $E_0(\theta)$ vs $\theta$ for $U=5$ and increasing strain values. Blue circles and red squares correspond to FRG and LCG, respectively. 
	In all cases, $E_0$ exhibits an oscillatory behavior with maxima and minima as $\theta$ varies from $0$ to $\pi$, as well as an overall increasing value for larger
	strains. The oscillations with strain direction are better defined
	for $\epsilon=0.30$ [Fig.~\ref{coupling_isotropicity_vs_GS_energies}(d)] but already visible for $\epsilon=0.15$ [see Fig.\ \ref{coupling_isotropicity_vs_GS_energies}(b)]. The minima occur for $\theta \simeq \pi/6$, $\pi/2$ and $5\pi/6$, which correspond to strain applied {\em orthogonal} to one of the lattice vectors ${\boldsymbol a}_j$.  In contrast, the $E_0$ maxima are observed around $\theta=0$, $\pi/3$ and $2\pi/3$, for which the strain direction lies {\em along} one of the lattice vectors. 
	The general behavior of $E_0$ increasing with $\epsilon$ can be seen to be associated with the suppression of antiferromagnetic correlations, which contribute to lowering the energy of the system. This is consistent with the fact that for LCG $E_0$ is lower, as additional singlet-like correlations appear across the edges. 
	We also notice that $E_0(\theta)$ is not fully symmetric around $\theta=\pi/2$, a direct consequence of the slanted geometry that defines the ends of the ribbon. Such asymmetry is even more pronounced for LCG, as seen in the red curves of Fig.~\ref{coupling_isotropicity_vs_GS_energies}. This feature becomes clear once one realizes that, for LCG, configurations (e) and (f) in Fig.~\ref{link_array} loop ${\boldsymbol a}_1$ lines onto themselves (see Fig.~\ref{schematic_ribbon}), in great contrast to what happens with the remaining ones: for example, ${\boldsymbol a}_2$ lines are alternatingly coupled with one another in (b) and (c).  
	
	To gain more intuition on the behavior of $E_0$ vs $\theta$ shown in Fig.~\ref{coupling_isotropicity_vs_GS_energies}, we analyze the nature of the maxima and minima of $E_0(\theta)$. Intuitively, one would expect the system to be more frustrated if all couplings were equal. It is thus reasonable to explore the distinct hoppings $t_j$ along different lattice vectors ${\boldsymbol a}_j$, as the strain changes them according to Eq.~\eqref{eq:strained_hoppings_main_text}. We quantify the isotropy of the couplings  by analyzing their distribution, as a smaller dispersion in hopping constants favors frustration in the system's ground state. The simplest measure
	of dispersion of hoppings is given by the variance $\sigma^2$. We introduce the inverse of  $\sigma^2$ as a measure of the homogeneity of the couplings,  defining a geometric frustration function as
	\begin{eqnarray}\label{coupling_isotropicity}
		\zeta(\theta)=\frac{1}{\sigma^2}=\frac{N_t(N_t-1)}{\sum_{j=1}^{N_t}\left(t_{ j } - \bar t\right)^{2}}
		=\frac{6}{\sum_{j=1}^{3}\left(t_{j}-\bar 
			t\right)^{2}},
	\end{eqnarray}
	where $\bar t=\left({t_{1}+t_{2}+t_{3}}\right)/3$, and $N_t=3$ is
	the number of distinct hoppings in the kagome lattice. Note that Eq.~\eqref{coupling_isotropicity}  
	depends implicitly on $\theta$ through both $t_j$ and $\bar t$.
	
	This quantity can be scaled and compared with the ground state energy $E_0(\theta)$ \footnote{The scaling factor is obtained as follows: with $\zeta_{max}\equiv \text{max} \left \{ \zeta ( \theta) \right \}$ and $\zeta_{min}\equiv \text{min} \left \{ \zeta ( \theta) \right \}$; the equivalent minimum (maximum) value of $E_0(\theta)$ for those angles in the FRG curve are $E_{min}$ ($E_{max}$). Then $\zeta \rightarrow \left ( \frac{\zeta -\zeta_{min}}{\zeta_{max}-\zeta_{min}}\right ) \left ( E_{max} - E_{min}\right) + E_{min}$ gives the desired scale for $\zeta (\theta)$ shown in Fig.~\ref{coupling_isotropicity_vs_GS_energies}.}.  Figure~\ref{coupling_isotropicity_vs_GS_energies} shows  $\zeta(\theta)$ as a dashed black curve; evidently, the scaled function $\zeta(\theta)$ follows $E_0(\theta)$ quite well. As mentioned above, the results for LCG are more affected by end effects, and the agreement is poorer. Nevertheless, the positions of maxima and minima predicted by $\zeta(\theta)$ agree quite nicely with $E_0(\theta)$ for both FRG/LCG configurations obtained via DMRG. These results suggest that the maxima of $E_0(\theta)$ are obtained whenever the geometric frustration in the system is higher, even for the case $U=5$ (metallic regime when unstrained) shown in Fig.~\ref{coupling_isotropicity_vs_GS_energies}.
	
	We now turn our attention to the effect of strain on the spin-spin correlations in the kagome ribbon.
	Figure~\ref{link_array} shows the heatmap of ${\cal C}_{\rm link}$ for angles where  $E_0$ and  $\zeta$ have maxima (i.e., higher frustration, $\theta=0$, $\pi/3$ and $2\pi/3$, left panels) or minima (lower frustration, $\theta=\pi/6$, $\pi/2$ and $5\pi/6$, right panels).   We note that  ${\cal C}_{\rm link}$ exhibits a structure of connected oblique Lieb-like lattices \cite{kagomeLiebDeformation} for strain directions where $\zeta$ is maximum (left panels).
	The links are homogeneous in strength/intensity throughout, with links stretched the most (those {\em along} $\theta$) having nearly vanishing correlation strength.
	In contrast, for angles where $E_0$ is minimum (right panels), the ${\cal C}_{\rm link}$ distribution appears as 
	quasi-independent 1D chains of strong correlation links at angles {\em orthogonal} to strain direction $\theta$. 
	
	These features in the overall link distribution are even more pronounced for larger values of the Coulomb interaction $U$, as charge fluctuations are suppressed and antiferromagnetic correlations enhanced overall.
	This is evident in Fig.~\ref{link_array_larger_U} that shows link-correlations for $U=10$ (upper) and $U=20$ (lower panels). Left and right panels refer to $\theta=\pi/3$ and $\pi/6$, which compare with Fig.~\ref{link_array}(c) and (b), respectively. Note that the Neel-like lines (right panels) and the oblique 2D lattice structures (left) are much more pronounced in this figure than in Fig.~\ref{link_array}.  We also note relatively little variation between $U=10$ and 20.
	
	While $\zeta (\theta)$ is a good indicator of $\theta$ values that most affect the ground state energy $E_0$, it does not directly quantify the spin frustration that may be present in the system. [Yet another indicator is discussed in Appendix~\ref{appendix_coupling_ratios}.
	A useful way to quantify the frustration content of the ground state is via the following quantity,
	\begin{eqnarray}\label{local_geometrical_saturation}
		f_j^G=1-\frac{4}{\tilde{J}_j}\left|\sum_{(m,n)\in\mathbb{C_B^{\text{j}}}}J_{mn}\left<S_m^z S_n^z\right>\right|,
	\end{eqnarray}
	which accounts for the spin-spin correlations within a given unit cell of the kagome system. In this expression, $J_{mn}=4t_{mn}^2/U$ represents the exchange coupling between neighbor spins $(m,n)$ within the unit cell considered, $\mathbb{C}_B^j$, and included in the sum. The unit cell used is shown by dashed magenta lines in Fig.~\ref{schematic_ribbon}.
	Finally, $\tilde{J}_j\equiv\sum_{(m,n)\in\mathbb{C_B^{\text{j}}}} \left | J_{mn} \right |$.
	In this definition, $f_j^G$ accounts for the couplings (including their the sign) between each pair of spins in all unit cells. Notice that $f_j^G \rightarrow 1$ for uncorrelated cells, while $f_j^G \rightarrow 0$ for both fully ferro- and antiferromagnetic correlated links.

\begin{figure}
		\hskip -0.5cm
		\includegraphics[scale=0.36]{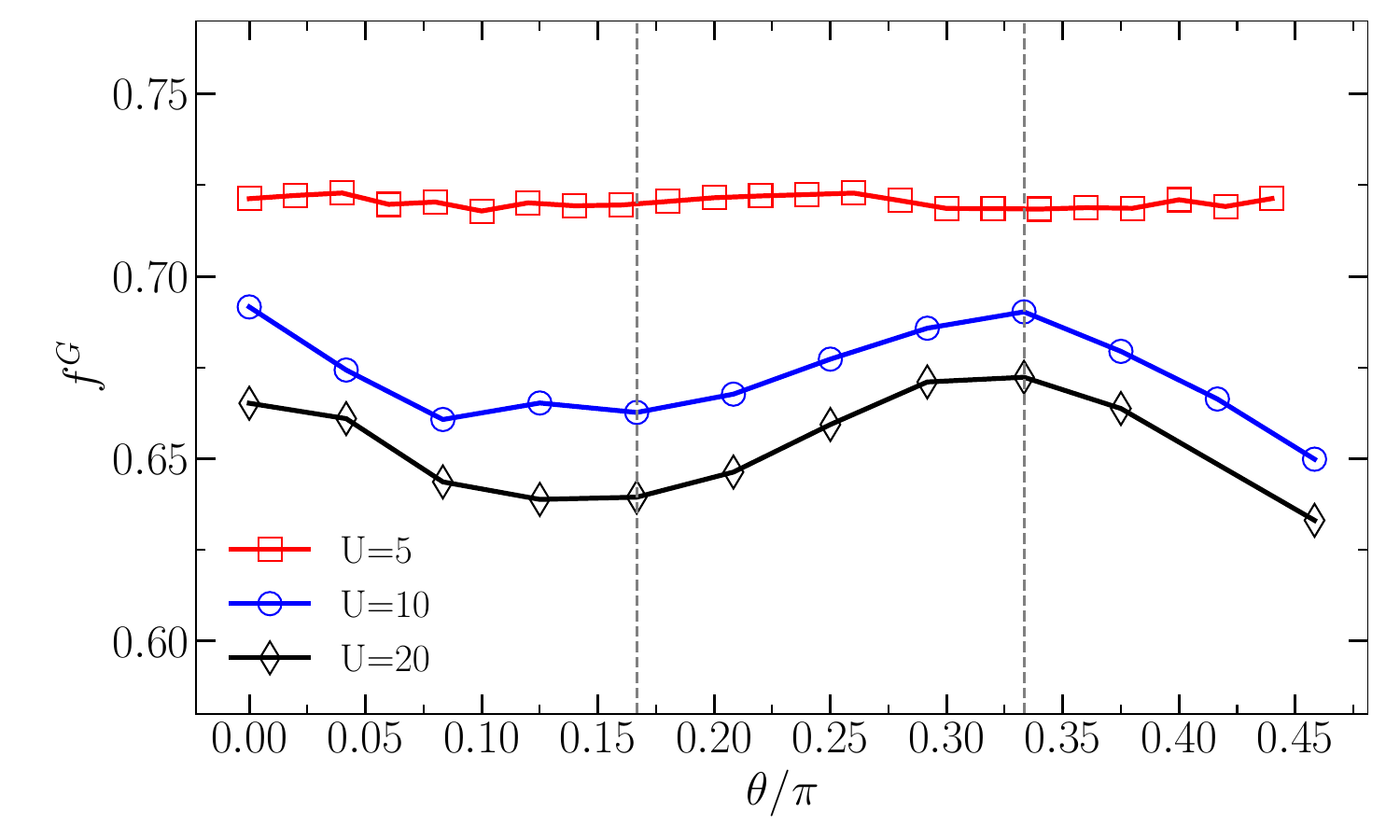}
		\caption{$f^G(\theta)$ vs $\theta$ for $\epsilon=0.30$, $U=5$ (red squares), $U=10$ (blue circles) and $U=20$ (black diamonds). Note that for larger values of $U$, the maxima and minima become more pronounced, as expected. Dashed lines indicate $\theta=\pi/6, \, \pi/3$, for the minimum and maximum in $E_0$.
		}
		\label{geometrical_saturation_U_comparison} 
	\end{figure}

	Figure~\ref{geometrical_saturation_U_comparison} shows the average of this indicator over the entire system, $ f^G(\theta)$, for three distinct values of $U$, over the range $0 < \theta< \pi/2$; apart from small end-effect asymmetry, similar results are obtained for $\pi/2< \theta< \pi$.  Notice that for large $U=10$ and 20, where antiferromagnetic correlations are better defined in the system, $f^G$ drops in value overall and exhibits well-pronounced maxima and minima that agree well with the corresponding behavior of $E_0(\theta)$ in Fig.~\ref{coupling_isotropicity_vs_GS_energies}.
	The nearly flat behavior of $ f^G(\theta)$ for $U=5$ is likely associated with charge fluctuations in the system, suppressing local moments in lattice sites for weak interactions.  This indicator is then not as effective in providing a quantitative assessment in the weak interaction regime, despite the clear impact of strain on the ${\cal C}_{\rm link}$ distribution. 
	\begin{figure*} 
		\centering
		\includegraphics[width=\textwidth]{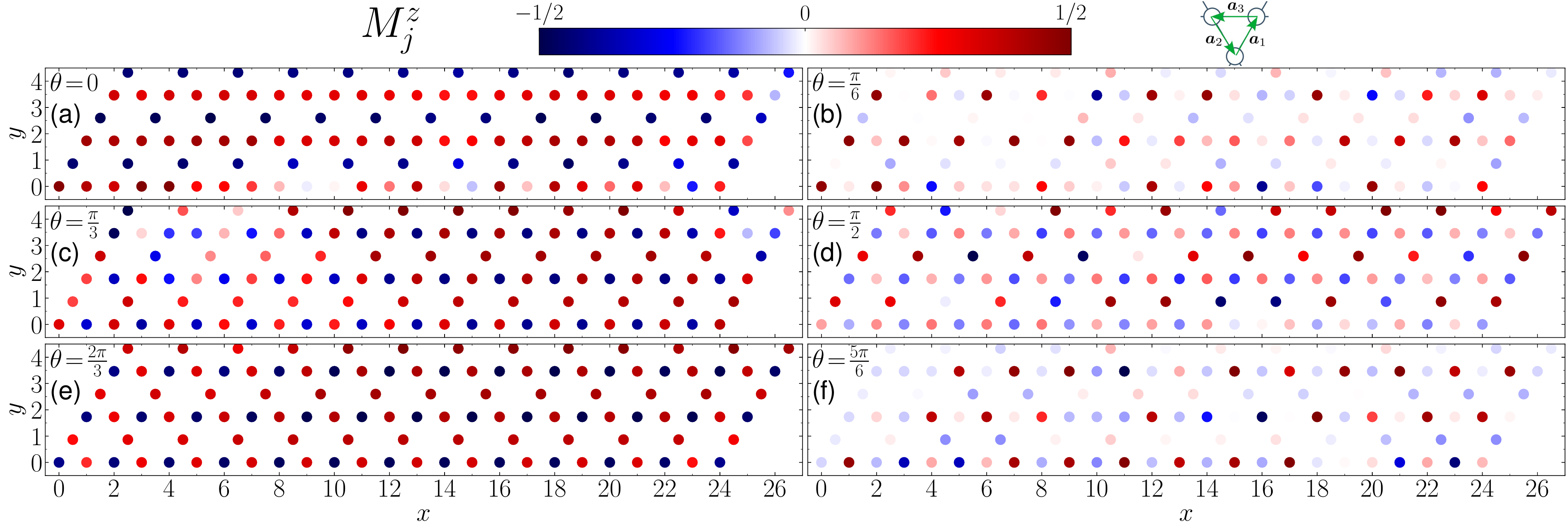}
		\caption{Magnetization $M^z_j$  for $V_z=0.05$ and $\theta=0$ (a), 
			$\theta={\pi}/{6}$ (b), $\theta={\pi}/{3}$ (c),  $\theta={\pi}/{2}$ (d), $\theta={2\pi}/{3}$ (e) and $\theta={5\pi}/{6}$ (f). For all panels, we set $U=5$ and $\epsilon=0.30$. Faint dots on the right panels \textit{(where $E_0$ has minima in Fig.~\ref{coupling_isotropicity_vs_GS_energies})} feature sites belonging to Neel-like lines, while dark red dots correspond to loose sites readily polarized by the field. Top-right inset helps identifying, e.g., the red lines in (c) as being along the strain direction $\boldsymbol{a_1}$.
			\label{local_magnetizations}}
	\end{figure*}
	\begin{figure*}
		\centering
		\includegraphics[width=\textwidth]{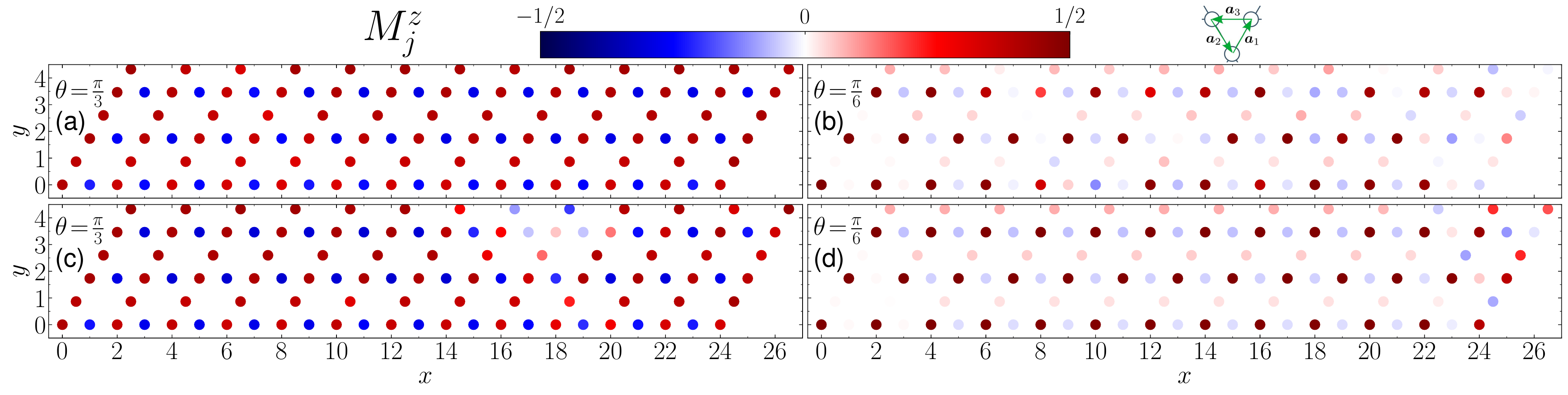}
		\caption{Local magnetization $M^z_j$ for $V_z=0.05$ and strain applied along $\theta=\pi/3$ (left)  and  $\theta=\pi/6$ (right). Upper and lower panels show results for $U=10$ and $U=20$, respectively.  As in  Fig.~\ref{local_magnetizations}, faint color dots on the right panels represent sites belonging to Neel AF lines, while dark red dots correspond to loose sites more easily polarized by the field. Notice how the most energetically favorable top-left to bottom-right red-then-blue pattern results along $a_2$-connected spins.
			For all pannels, $\epsilon=0.30$.
			\label{local_magnetizations_large_U}
		}
	\end{figure*}
	% 
	%
	%\begin{figure}[t!]
	\begin{figure}
		\includegraphics[scale=0.48]{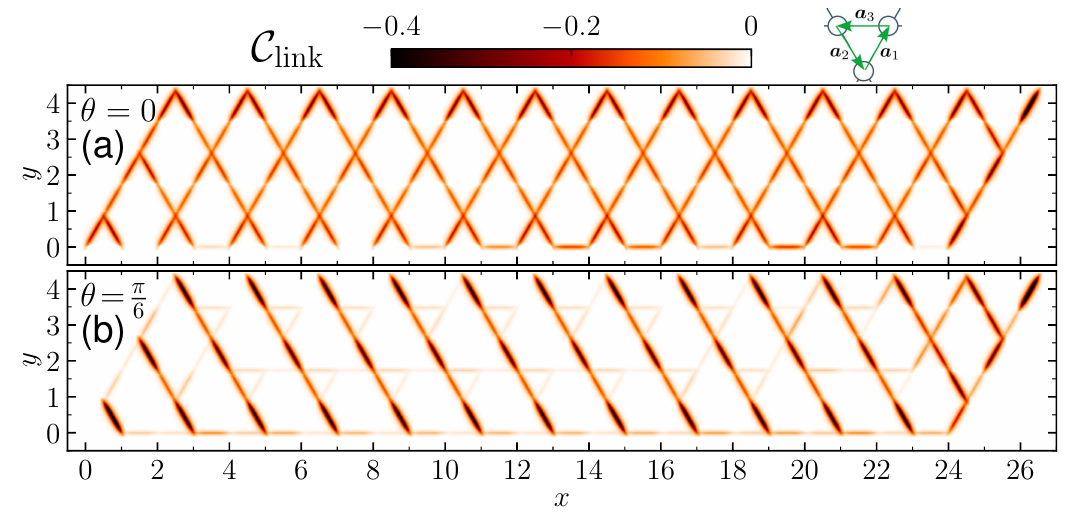}
		\caption{Link correlations for $\epsilon=0.30$, $V_z=0.05$, for $\theta=0$ (a), and $\theta=\pi/6$ (b), corresponding to maximum and minimum of the frustration quantifier shown in Fig.~\ref{coupling_isotropicity_vs_GS_energies}. In panel (a) all sites are connected with neighbors along Neel AF lines forming 2D lattice. In panel (b) Neel lines are disconnected from each other, leaving many sites nearly fully uncorrelated. In both panels $U=5$.
			\label{link_correls_first_stationary_energies__V_005}}
	\end{figure}

	\subsection{Coexistent magnetization and Neel structures}
	\label{magnetic_ordering_coexistence_subsection}
	Let us go back to link correlations for the strained case and take a closer look at Fig.~\ref{link_array}.  Note that for angles where $\zeta$ or $E_0$ have a minimum (right panels of Fig.~\ref{link_array}), the formation of 1D correlated lines conspire to leave the sites between lines only loosely coupled to their neighbors. This suggests the coexistence of antiferromagnetic links and paramagnetic sites in the lattice.  One would further expect these two types of sites (free and correlated) would respond differently to an applied magnetic field. The (antiferromagnetically) correlated spins would require a larger field to be polarized, as the field should compete with the effective exchange interaction. On the other hand, the loose spins would align readily with the field.

	To verify this reasoning, we apply a small magnetic field along the 
	$z$-direction to the entire system and analyze the behavior 
	of the spatially-resolved magnetization  over the lattice. 
	Figure~\ref{local_magnetizations} shows the local $z$-magnetization $M_j^z=\langle S_j^z\rangle$ for the ground state in a small applied field $V_z=0.05$ and for the same strain angles as in Fig.~\ref{link_array}. Colorful dots at each site of the lattice indicate whether that spin is polarized towards $+\hat{\boldsymbol z}$ (red), $-\hat{\boldsymbol z}$ (blue) or unpolarized (white), as per the color scale. Let us analyze what is shown in Fig.~\ref{local_magnetizations}. Starting with the case of $\theta=0$ [Fig.~\ref{local_magnetizations}(a)] we observe a pattern consisting of alternating positive/negative $M_j^z$ for $j$ on the line along the vectors ${\boldsymbol a}_1$ or ${\boldsymbol a}_2$  with all positive $M_j^z$ along ${\boldsymbol a}_3$. This pattern can be connected with Fig.~\ref{link_array}(a) as follows: since the weakest link correlations are seen there along ${\boldsymbol a}_3$, the application of a magnetic field easily polarizes all spins parallel to the field for sites along ${\boldsymbol a}_3$; in contrast, sites along ${\boldsymbol a}_1$ and ${\boldsymbol a}_2$ align antiferromagnetically to those polarized by the field. The AF links remain intact because, for this weak field, it is still energetically favorable to form the Neel lines along with strong link correlations. Similar conclusions can be made for the cases of $\theta=\pi/3$ and $\theta=2\pi/3$, except that the weakest links in these cases lie along ${\boldsymbol a}_1$ and ${\boldsymbol a}_2$, respectively. We emphasize that sites aligned with the field and lying along ${\boldsymbol a}_3$ in Fig.~\ref{local_magnetizations}(a) are consistent with the formation of AF lines along ${\boldsymbol a}_1$ and ${\boldsymbol a}_2$ in Fig.~\ref{link_array}(a), and that the AF many-body correlations are not suppressed for small fields, as one would expect. 
	
	Let us now look at the right panels of Fig.~\ref{local_magnetizations} for which the frustration $\zeta$ and $E_0$ have minima. The Neel lines with strong link correlation of Fig.~\ref{link_array}(b) (for $\theta=\pi/6$) lie parallel to the lattice vector ${\boldsymbol a}_2$. This is manifested in the faint, nearly unpolarized sites along ${\boldsymbol a}_2$ in Fig.~\ref{local_magnetizations}(b), suggesting that the Neel lines are not affected by the external field. Note also in  Fig.~\ref{local_magnetizations}(b) that the alternating polarization of sites are not as well pronounced as for the maximal $E_0$ cases.
	Most important, however, are the dark red sites between Neel lines which strongly polarize regardless of their neighbors, and are scattered throughout the system. This is consistent with the idea that loosely connected sites exist over the lattice that respond efficiently to the applied field.
	
	The coexistence of distinct magnetic structures, strongly correlated links and easily polarizable loose sites  is expected to be more pronounced for larger values of  $U$. To test this, Fig.~\ref{local_magnetizations_large_U} shows $M_j^z$ for $U=10$ (upper row) and $U=20$ (lower row), for the same magnetic field $V_z=0.05$. Left and right panels correspond to $\theta=\pi/3$ and $\theta=\pi/6$, respectively, corresponding to the lattice strains in Fig.~\ref{link_array_larger_U}.
	We again observe on the left panels that the weakly correlated sites along the $\theta$ direction (${\boldsymbol a}_1$) become easily polarized by the field, while the neighboring sites are/remain strongly AF correlated along ${\boldsymbol a}_2$ and ${\boldsymbol a}_3$ [as in Fig.~\ref{local_magnetizations}(c)].  The right panels in Fig.~\ref{local_magnetizations_large_U} show that the energy minimizes
	on alternating diagonal lines that weakly polarize in the field while coexisting with strongly polarized sites through the structure, similarly to Fig.~\ref{local_magnetizations}(b). 
	
	%\rafael{In a first glance, one might think that the spin pattern (right-blue from top to bottom) would have a red $\leftrightarrow$ blue symmetry along Neel lines, but the presence of polarized interstitial sites guarantee non-degeneracy as follows. Should the blue and red sites be flipped, it is easy to see that the decisive energy balances are due to whatever results along weak links, since it is not the true many-body ("connected") correlations that define energies in the Heisenberg model. Hence, the $a_3$-connected links would all be ferromagnetically aligned, whilst $a_1$ ones would be antiferro, thus reversing what is displayed in Fig.~\ref{local_magnetizations_large_U} (b) and (d). The state is then a higher energy one since there are more $a_3$ than $a_1$-connected links. \textbf{QUESTION: A conjecture posits itself: could in the bulk material ($L_x, L_y>>1$) could the ground state be degenerate...?}}
	
	Finally, one may wonder whether the applied magnetic field destroys the many-body spin correlations. We expect them to be robust along well defined Neel lines. To confirm this, Fig.~\ref{link_correls_first_stationary_energies__V_005} shows the link
	correlations for systems as in Fig.~\ref{link_array}(a) and (b) {\em under an applied field}.  Comparing these two figures, it is clear that correlations along the strong links are indeed preserved for the applied field $V_z=0.05$. Interestingly, correlations along the lower edge are more affected, likely due to their fewer neighbors. 
	Notice again that the Neel lines are parallel to the lattice vector ${\boldsymbol a}_2$ in Fig.~\ref{link_correls_first_stationary_energies__V_005}(b) and \ref{link_array}(b) as the sites in between adjacent lines are weakly correlated to their neighbors and are then strongly polarizable by the field, as seen in Fig.~\ref{local_magnetizations}(b). These sites behave as a collection of paramagnetic (noninteracting) polarizable spins, while the sites belonging to Neel lines form a 	separate set of AF 1D chains.  The coexistence of these two classes of sites and the dependence of their appearance and orientation with applied strain represent the main unexpected results of this work. 
	%\rafael{Loose sites have easily broken correlations, which are traded away under the possibility of field alignment. The "spurious" dark links, appearing even where correlations were not expected on a hopping-strenght basis (as seen in Figs.~\ref{link_corr} and ~\ref{link_array_larger_U}), occur on edges that don't participate in the construction of triangles. Apart from arising between sites with fewer neighbors, this "correlation concentration" might indicate weak entanglement that locally suffers a "phase transition", if we may abuse language like that: the applied field is strong enough to (at least partially) break the singlet-like configuration in favor of polarizing the spins.}
	
	\section{Conclusions}
	We have studied a kagome nanoribbon using a Hubbard model to describe how anisotropies induced by uniaxial strain break the symmetry and affect the 
	correlations due to electronic repulsion.  We find that even for relatively weak interactions, the competing correlations and frustration develop coexisting systems of strongly antiferromagnetic correlations between neighboring sites and weakly connected sites that are easily polarizable in a magnetic field.  The spatial distribution of the strong links and interspersed paramagnetic sites depends strongly on the orientation of the uniaxial strain, and opens the possibility of exploring an experimentally tunable probe that can give rise to different correlated behavior in a given lattice.
	The need to consider charge fluctuations to define a function that quantifies correlations, especially for weak $U$ values, and that complements $f^G$, remains an interesting theoretical question. 
	With the growing interest in metallic kagome materials displaying strongly correlated behavior, it may be possible to apply external strain fields and explore some of the phenomena our models describe.  It would be interesting, moreover, to explore how distinct Fermi levels and proximity to van Hove singularities in different materials affect this behavior. As a final reflection, there might be a connection between our results and the spin Jahn-Teller effect. The idea that spontaneous distortions could play a role in kagome materials is appealing and deserves future investigation.
	\label{conclusions}

	\begin{acknowledgments}
		EV acknowledges financial support from CNPq (Process 311366/2021-0), FAPEMIG (Process PPM-00631-17) and CAPES under UFU-CAPES Print initiative.
		SEU acknowledges support by the US Department of Energy, Office of Basic Energy Sciences, Materials Science and Engineering Division.
	\end{acknowledgments}
	
	\appendix
	\section{Strained couplings}
	\label{appendix_strained_couplings}
	The deformed nearest-neighbor vectors are given by%
	\begin{eqnarray}
		\boldsymbol{a'}_{j}=\left(\boldsymbol{I}+\overline{\epsilon}\right)\boldsymbol{a}_{j}, \label{eq:strain_deformed_vecs}
	\end{eqnarray}
	with 
	\begin{eqnarray*}
		\boldsymbol{a}_{1}&=&\frac{a_{0}}{2}\left(1,\sqrt{3}\right),\quad
		\boldsymbol{a}_{2}=\frac{a_{0}}{2}\left(1,-\sqrt{3}\right), \quad
		\boldsymbol{a}_{3}=a_{0}\left(-1,0\right),
	\end{eqnarray*}
	and the strain tensor $\overline{\epsilon}$ is given by Eq.~\eqref{eq:strain-tensor-main-text}. We set the lattice constant $a_0=1$.
	Using trigonometric identities one can readily write,
	\begin{eqnarray}
		\overline{\epsilon}=\frac{\epsilon}{2}\left\{ \left(1-\nu\right)\boldsymbol{I}+\left(1+\nu\right)\tilde{\text{R}}\left(\theta\right)\right\} ,\label{eq:decomposed_strain_tensor}
	\end{eqnarray}
	where 
	\begin{eqnarray}
		\tilde{\text{R}}\left(\theta\right)=\left[\begin{array}{cc}
			\cos\left(2\theta\right) & \sin\left(2\theta\right)\\
			\sin\left(2\theta\right) & -\cos\left(2\theta\right)
		\end{array}\right]=\tilde{\text{R}}^{\text{T}}\left(\theta\right),\label{eq:skew_rotation_of_strain_tensor}
	\end{eqnarray}
	which is then an orthogonal symmetric tensor: 
	\begin{eqnarray}
		\tilde{\text{R}}\left(\theta\right)\tilde{\text{R}}^{\text{T}}\left(\theta\right)=\tilde{\text{R}}^{\text{T}}\left(\theta\right)\tilde{\text{R}}\left(\theta\right)=\left[\tilde{\text{R}}\left(\theta\right)\right]^{2}=\boldsymbol{I}.\label{eq:skew_rotation_orthogosymmetry}
	\end{eqnarray}
	Now, by defining $\alpha_{\pm}\equiv \epsilon \left(1\pm\nu\right)/2$,
	%\begin{equation}
	%\alpha_{\pm}\equiv\frac{\left(1\pm\nu\right)}2}\epsilon\label{eq:alpha_plus_or_minus}
%\end{equation}
we can recast Eq.~\eqref{eq:strain_deformed_vecs} as
\begin{eqnarray}
	\boldsymbol{a'_{j}}\left(\theta\right)=\left[\left(1+\alpha_{-}\right)\boldsymbol{I}+\alpha_{+}\tilde{\text{R}}\left(\theta\right)\right]\boldsymbol{a_{j}}.\label{eq:decomposed_form_strained_vecs}
\end{eqnarray}

The strained hopping terms to be calculated are given by
\begin{equation}
	t_{j}\left(\theta\right)=t_{j}\exp\left\{ -\beta\left[\left\Vert \boldsymbol{a'_{j}}\left(\theta\right)\right\Vert -1\right]\right\} .\label{eq:STRAINED_HOPPINGS}
\end{equation}

The strained vectors moduli $\left\Vert \boldsymbol{a'_{j}}\left(\theta\right)\right\Vert =\sqrt{\boldsymbol{a'_{j}}\left(\theta\right){\cdot}\boldsymbol{a'_{j}}\left(\theta\right)}$, 
and since $\tilde{\text{R}}$ is both symmetric and orthogonal,
we get
\begin{equation}
	\boldsymbol{a'_{j}}\left(\theta\right){\cdot}\boldsymbol{a'_{j}}\left(\theta\right)\!=\!\left[\alpha_{+}^{2}+\left(1+\alpha_{-}\right)^{2}\right]\boldsymbol{\boldsymbol{a_{j}}{\cdot}\boldsymbol{a_{j}}}+2\alpha_{+}\left(1+\alpha_{-}\right)\boldsymbol{a_{j}}{\cdot}\boldsymbol{\tilde{a}_{j}\left(\theta\right)},\label{eq:strained_vecs_squared_modulus}
\end{equation}
where $\boldsymbol{\tilde{a}_{j}\left(\theta\right)}\equiv\tilde{\text{R}}\left(\theta\right)\boldsymbol{a_{j}}.$ 

Finally, using the traditional definition of antiferromagnetic energy scale $J_{n}=4t_{n}^{2}/U$ we obtain the effective strain-induced Heisenberg couplings,
%\textbf{\rafael{QUESTION: Would ~\ref{eq:STRAINED_HOPPINGS} be the expression for $J_n(\theta)$ should we have started directly with a Heisenberg model, instead? I don't know if we need to discuss that...}}
%
\begin{widetext}
	\begin{eqnarray}
		J_{n}\left(\theta\right)=J_{n}\exp\left\{ -2\beta\left[\left(\left[\alpha_{+}^{2}+\left(1+\alpha_{-}\right)^{2}\right]+2\alpha_{+}\left(1+\alpha_{-}\right)\frac{1}{\text{x}_{n}^{2}+\text{y}_{n}^{2}}\left[\left(\text{x}_{n}^{2}-\text{y}_{n}^{2}\right)\cos\left(2\theta\right)+2\text{x}_{n}\text{\ensuremath{\text{y}_{n}}}\sin\left(2\theta\right)\right]\right)^{1/2}-1\right]\right\}. \label{eq:strained_heisenberg_couplings}
	\end{eqnarray}
\end{widetext}
where $\boldsymbol{a_{j}}=(\text{x}_{j}, \text{y}_{j})$.

\section{Coupling ratios}
\label{appendix_coupling_ratios}
\begin{figure}
	\centering
%	\hskip -0.95cm
	\includegraphics[scale=0.58]{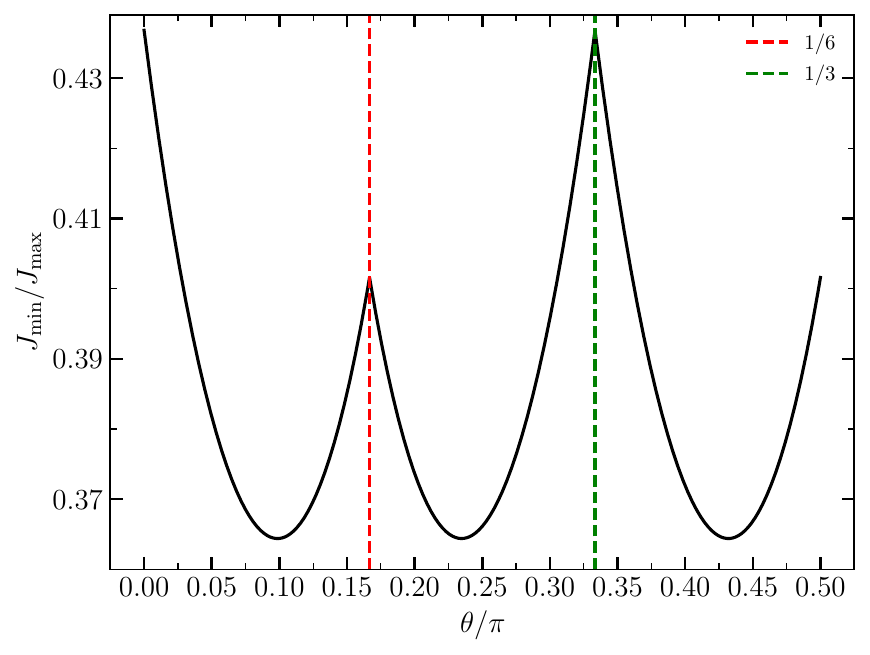}
	\caption{$r(\theta)$ vs $\theta$ for $\epsilon=0.30$. Note that this quantity is U-independent. Dashed lines indicate $\theta=\pi/6, \, \pi/3$, for a minimum and maximum in $E_0$.
	}
	\label{heisenberg_coupling_ratios} 
\end{figure}
For each angle $\theta$, we display the ratio $r(\theta)\equiv J_{\text{min}}(\theta)/J_{\text{max}}(\theta)$ between the smallest and largest couplings in Fig.~\ref{heisenberg_coupling_ratios}. This is a non-analytical function, with cusps located at the stationary points in Fig.~\ref{coupling_isotropicity_vs_GS_energies}. Consistent with Eqs.~\eqref{coupling_isotropicity} and \eqref{local_geometrical_saturation} the plot reflects that $r(\theta)$ values closer to one indicate greater homogeneity of couplings, thus implying higher frustration--hence the peak at, e.g., $\theta=\pi/3$ being taller than the one at $\theta=\pi/6$. 
We conclude that $r(\theta)$ is a much coarser way to infer frustration, and therefore not as useful indicator as those discussed in the main text.

% \bibliographystyle{plain}
%\bibliography{references}

%apsrev4-2.bst 2019-01-14 (MD) hand-edited version of apsrev4-1.bst
%Control: key (0)
%Control: author (8) initials jnrlst
%Control: editor formatted (1) identically to author
%Control: production of article title (0) allowed
%Control: page (0) single
%Control: year (1) truncated
%Control: production of eprint (0) enabled
%

\end{document}